\documentclass[12pt,preprint]{aastex}

\usepackage{ulem} 

\usepackage{longtable}
\usepackage{rotating}

\slugcomment{{\Large{{\it Research Note}}}}

\shorttitle{Astrometry and Photometry with {\it HST}-WFC3/UVIS. I.}
\shortauthors{Bellini, A. \& Bedin, R.~L.}

\begin{document}

\title{Astrometry and photometry with {\it HST}-WFC3. I. ~~~~~~~~~~
Geometric distortion corrections of F225W, F275W, F336W bands of the
UVIS-channel.}

\author{A. Bellini\altaffilmark{1}} 
\affil{Dipartimento di Astronomia, Universit\`a di Padova, 
Vicolo dell'Osservatorio 3, 35122 Padova, Italy}
\email{bellini@stsci.edu, [andrea.bellini@unipd.it]}

\author{L.~R. Bedin}
\affil{Space Telescope Science Institute, 3700 San Martin Drive, 
Baltimore, MD 21218, USA}
\email{bedin@stsci.edu}

\altaffiltext{1}{Visiting PhD Student at STScI under the {\it ``2008
graduate research assistantship''} program.}

\date{\footnotesize{\it ~~~~~~~~~~~~~~~~~~~~~~~Received 2009 September
23; Accepted 2009 October 16}}

\begin{abstract}
An accurate geometric distortion solution for the \textit{Hubble Space
Telescope} UVIS-channel of Wide Field Camera 3 is the first step
towards its use for high precision astrometry. In this work we present
an average correction that enables a relative astrometric accuracy of
$\sim$1 mas (in each axis for well exposed stars) in three broad-band
ultraviolet filters (F225W, F275W, and F336W).  More data and a better
understanding of the instrument are required to constrain the solution
to a higher level of accuracy.
\end{abstract}

~\\ 
\newpage
%#####################################################################
%#####################################################################
\section{Introduction, Data set, Measurements}
\label{sec1}

The accuracy and the stability of the geometric distortion\footnote{ A
specification is needed. With the term ``geometric distortion'', or
GD, which we will use hereafter, we are lumping together several
effects under the same term: the optical field-angle distortion
introduced by camera optics, light-path deviations caused by the
filters, non-flat CCDs, alignment errors of CCDs on the focal plane,
etc..}  (GD) correction of an instrument is at the basis of its use
for high precision astrometry.  The particularly advantageous
conditions of the {\it Hubble Space Telescope (HST)} observatory make
it ideal for imaging-astrometry of (faint) point sources.  The
point-spread functions (PSFs) are not only sharp and (essentially)
close to the diffraction limit --which directly results in high
precision positioning-- but also the observations are not plagued by
atmospheric effects (such as differential refraction, image motion,
differential chromatic refraction, etc), which severely limit
ground-based astrometry.  In addition to this, {\it HST} observations
do not suffer from gravity-induced flexures on the structures of the
telescope (and camera), which add (relatively) large instabilities in
the GD of ground-based images, and make its corrections more
uncertain.

Last May 14, the brand-new {\it Wide Field Camera 3} (WFC3) was
successfully installed during the {\it Hubble Servicing Mission 4}
(SM4, May 12-24 2009).  After a period of intense testing,
fine-tuning, and basic calibration, last September 9th, 2009, the
first calibration- and science-demonstration images were finally made
public.

Our group is active in bringing {\it HST} to the {\it state of the
art} of its astrometric capabilities, that we used for a number of
scientific applications (e.g.\ from King et al.\ 1998, to Bedin et
al.\ 2009, first and last accepted papers).  Now that the {\it
``old''} ACS/WFC is successfully repaired, and that the new
instruments are installed, our first step is to extend our astrometric
tools to the new instruments (and to monitor the old ones).  This
paper is focused on the geometric distortion correction of the
\textit{UV/Optical} \textit{(UVIS)} channel of the WFC3.  Since the
results of these efforts might have some immediate public utility
(e.g.\ relative astrometry in general, stacking of images,
UV-identification of X-counterparts such as pulsars and CVs in
globular clusters, etc.), we made our results available to the
WFC3/UVIS user-community.

We immediately focused our attention on a deep UV-survey of the core
of the Galactic globular cluster $\omega$~Centauri (NGC~5139), where
some well dithered images were collected.  The dense --and relatively
flat-- stellar field makes the calibration particularly suitable for
deriving and monitoring the GD on a relatively small spatial scale.
In addition, while most of the efforts to derive a GD correction will
be concentrated on relatively redder filters, we undertook a study to
determine the GD solutions of the three bluest broad-band filters
(with the exception of F218W): F225W, F275, and F336W.

The WFC3/UVIS layout is almost indistinguishable from that of
ACS/WFC\footnote{\sf http://www.stsci.edu/hst/wfc3}:\ two E2V thinned,
backside illuminated and UV optimized 2k$\times$4k CCDs contiguous on
the long side of the chip, and covering a field of view (FoV) of
$\sim160$$\times$$160$ arcsec$^2$.  The $\omega$~Cen data set used
here consists of 9$\times$350~s exposures in each of the filters
F225W, F275W, and F336W. The nine images follow a squared 3$\times$3
dither-pattern with a step of about 40 arcsec (i.e.\ $\sim$1000
pixels), and were all collected on July 15, 2009.

We downloaded the standard pipe-line reduced {\sf FLT} files from the
archive. The {\sf FLT} images are de-biased and flat-field corrected,
but {\it no} pixel-resampling is performed on them.  The {\sf FLT}
files are multi-extension fits (MEF) on which the first slot contains
the image of what --hereafter-- we will call chip 1 (or simply
[1]). The second chip, instead, is stored in the fourth slot of the
MEF, and we will refer to it as chip 2 (or [2]).  [Note that others
might choose a different notation].  Our GD corrections refer to the
raw pixel coordinates of these images.

\begin{figure*}[t!]
\centering
\includegraphics[width=14cm]{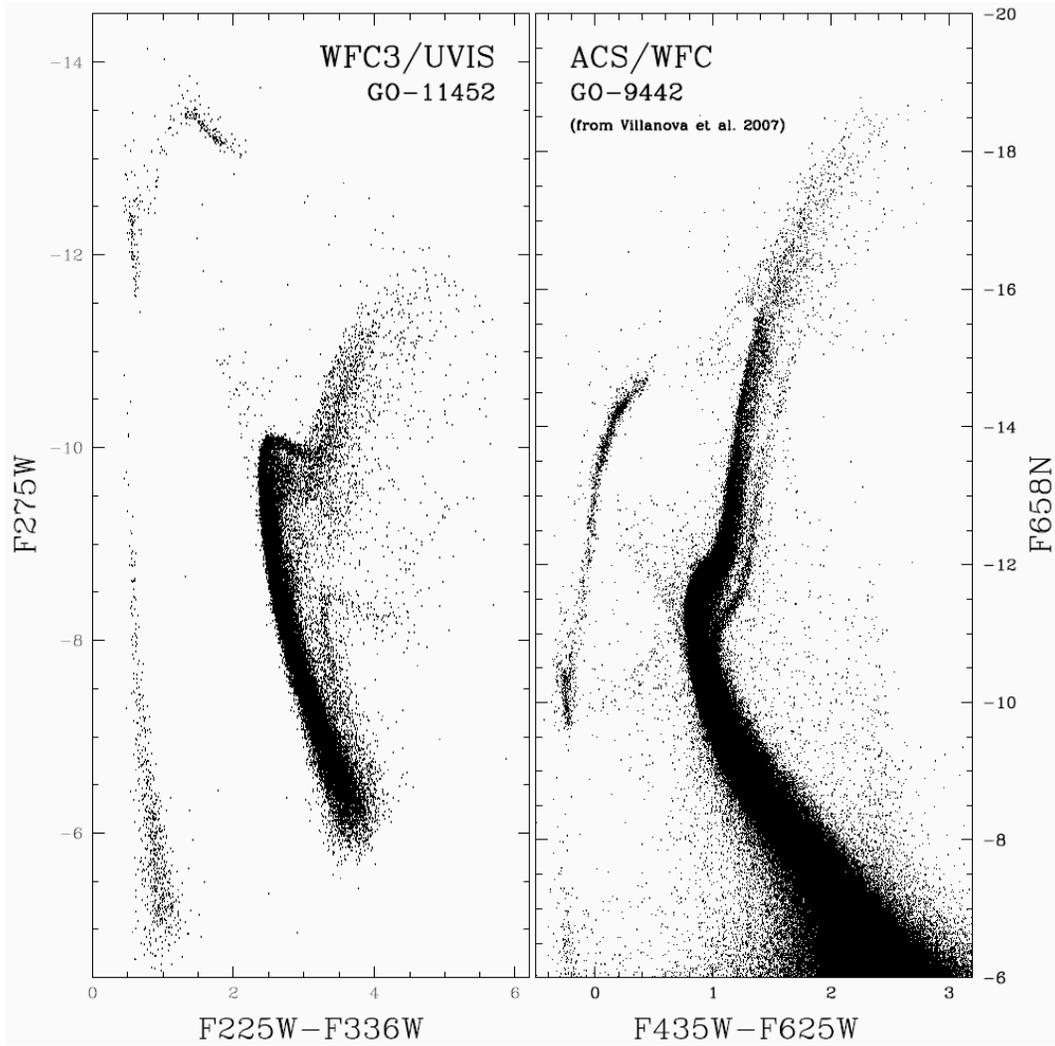}
\caption{ {\it Left:} Preliminary color-magnitude diagram of the
bright stars in the new WFC3/UVIS data set (fluxes are neither
pixel-area- nor L-flat-corrected).  {\it Right:} Color-magnitude
diagram of the stars in our ACS/WFC master frame (from Villanova et
al.\ 2007).  Both plots are in instrumental magnitudes.  }
\label{fig:0A}
\end{figure*}

The fluxes and positions were obtained from a code mostly based on the
software {\sf img2xym\_WFI} by Anderson et al.\ (2006).  This is
essentially a spatially variable PSF-fitting method.  We were pleased
to see that for the WFC3/UVIS images of this data set the PSFs were
only marginally undersampled.  Left panel of Figure~\ref{fig:0A} shows
a preliminary color-magnitude diagram in the three filters for the
bright stars in the WFC3/UVIS data set.  In a future paper of this
series we will discuss the PSF, its spatial variation and stability,
as well as L-flats\footnote{Residual low-frequency flat-field
structure (L-flat) cannot be accurately determined from ground-based
calibration data or internal lamp exposures.  L-flats need to be
determined from on-orbit science data, for example from multiple
observations of stellar fields with different pointings and roll
angles (van der Marel 2003).}, pixel-area corrections, and recipes for
deep photometry in stacked-images.

\begin{figure*}[t!]
\centering
\includegraphics[width=14cm]{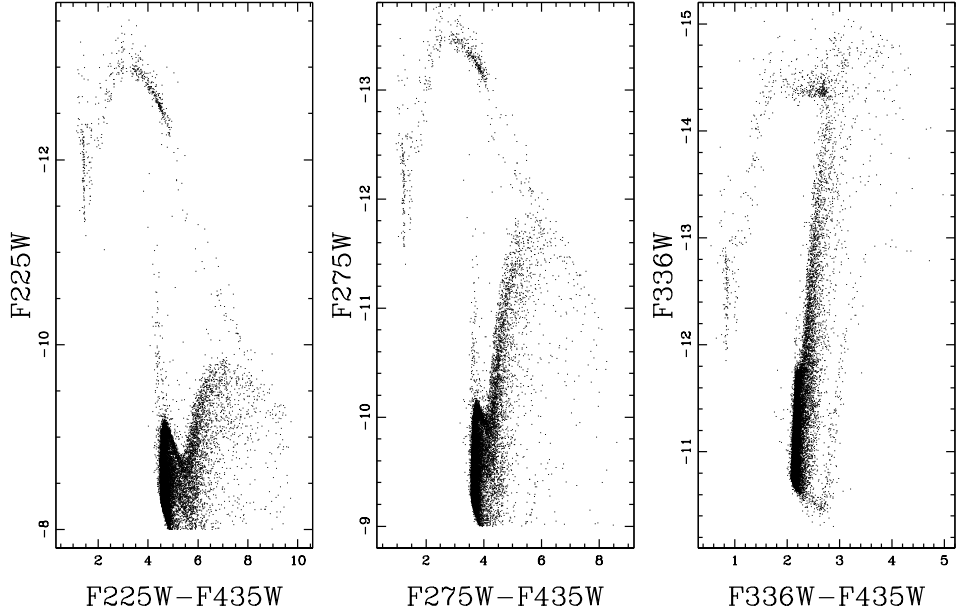}
\caption{{\it Left:} Color-magnitude diagrams of the high S/N stars in
common with the ACS/WFC master frame (F435W), actually used to derive
the geometric distortion correction, for each of the WFC3/UVIS filters
(F225W, F275W, and F336W). }
\label{fig:0B}
\end{figure*}

~\\
\newpage
~\\
\newpage
~\\
\newpage

\section{The Geometric Distortion Solutions}
\label{sec2}

The most straightforward way to solve for the GD would be to observe a
field where there is a priori knowledge of the positions of all the
stars in a distortion-free reference frame.  [A distortion-free
reference frame is a system that can be transformed into any other
distortion-free frame by means of {\it conformal
transformations\footnote{A conformal transformation between two
catalogs of positions is a four-parameter linear transformation,
specifically: rigid shifts in the two coordinates, one rotation, and
one change of scale, i.e. the shape is preserved.  }.}]  Geometric
distortion would then show itself immediately as the residuals between
the observed relative positions of stars and the ones predicted by the
distortion-free frame (on the basis of a conformal transformation).

Thankfully, we possess such an {``astrometric flat-field''}, moreover
with the right magnitude interval, source density, and accuracy.  This
reference frame is the mosaic of 3$\times$3 {\it Wide Field Channel
(WFC)} of the {\it Advanced Camera for Surveys (ACS)} fields collected
-- at the end of June 2002-- under the program GO-9442 (PI:\ Cool)
reduced by Jay Anderson and published in Villanova et al.\ (2007).
This reference frame was obtained from a total of 18 short and 90 long
ACS/WFC exposures in the filters: F435W, F625W, and F658N (see
Villanova et al.\ 2007 for details).  The entire field covers an area
of 10$\times$10 arcmin$^2$, and can be considered distortion-free at
the $\sim$0.5 mas level. The catalog contains more than 2 million
sources, and we will refer to it as {\it master frame}, and to the
coordinates of each $i$-source in it with the notation ($X^{\rm
master}_i$,$Y^{\rm master}_i$).  A color-magnitude diagram for the
stars in the master frame is shown on the right panel of
Figure~\ref{fig:0A}.

To derive the WFC3/UVIS GD corrections we closely follow the
procedures described in Anderson \& King (2003, hereafter AK03) used
to correct the GD for each of the four detectors of WFPC2.  We
represent our solution with a third-order polynomial, which is able to
provide our final GD correction to the $\sim$0.025 pixel level in each
coordinate ($\sim$1 mas). Higher orders proved to be unnecessary.

Having a separate solution for each chip, rather than one that uses a
common center of the distortion in the FoV, allows a better handle of
potential individual detector effects (such as a different relative
tilt of the chip surfaces, relative motions, etc.).  We adopted as the
center of our solution, for each chip, the point
$(x_\circ,y_\circ)_{k=1,2}=(2048,1025)$ [in the raw pixel coordinates,
to which we will refer hereafter as ($x,y$)].

For each $i$-star of the master list, in each $k$-chip of each
$j$-MEF-file, the distortion {\it corrected} position ($x^{\rm
corr},y^{\rm corr}$) is the observed position $(x,y)$ plus the
distortion correction $(\delta x,\delta y)$:
$$
\left\{
\begin{array}{c}
\displaystyle x_{i,j,k}^{\rm corr}=x_{i,j,k}+\delta x_{i,j,k}(\tilde{x}_{i,j,k},\tilde{y}_{i,j,k})\\
\displaystyle y_{i,j,k}^{\rm corr}=y_{i,j,k}+\delta y_{i,j,k}(\tilde{x}_{i,j,k},\tilde{y}_{i,j,k}),\\
\end{array}
\right.
$$ where $\tilde{x}_{i,j,k}$ and $\tilde{y}_{i,j,k}$ are the
normalized positions, defined as:
$$
\left\{
\begin{array}{c}
\displaystyle\tilde{x}_{i,j,k}= \frac{x_{i,j,k}-(x_\circ)_{k}}{(x_\circ)_{k}}\\
\displaystyle\tilde{y}_{i,j,k}= \frac{y_{i,j,k}-(y_\circ)_{k}}{(y_\circ)_{k}}.\\
\end{array}
\right.
$$ Normalized positions make it easier to recognize the magnitude of
the contribution given by each solution term, and their numerical
round-off.

The final GD correction for each star, in each chip/image, is given by
these two third-order polynomials (we omitted here $i,j,k$ indexes for
simplicity):
$$
\left\{
\begin{array}{rcl}
\displaystyle     \delta    x&\!\!\!=\!\!\!&a_1\tilde{x}+a_2\tilde{y}+
a_3\tilde{x}^2    +   a_4    \tilde{x}\tilde{y}+    a_5\tilde{y}^2   +
a_6\tilde{x}^3+a_7\tilde{x}^2\tilde{y}   +  a_8\tilde{x}\tilde{y}^2  +
a_9\tilde{y}^3\\                  \displaystyle                 \delta
y&\!\!\!=\!\!\!&b_1\tilde{x}+b_2\tilde{y}+    b_3\tilde{x}^2   +   b_4
\tilde{x}\tilde{y}+                  b_5\tilde{y}^2                  +
b_6\tilde{x}^3+b_7\tilde{x}^2\tilde{y}   +  b_8\tilde{x}\tilde{y}^2  +
b_9\tilde{y}^3.\\
\end{array}
\right.
$$

Our GD solution is thus fully characterized by 18 coefficients:
$a_1,\dots, a_9$, $b_1,\dots, b_9$.  However, as done in AK03, we
constrained the solution so that, at the center of the chip, it will
have its $x$-scale equal to the one at the location
$(x_\circ,y_\circ)$, and the corrected axis $y^{\rm corr}$ has to be
aligned with its $y$-axis at the location $(x_\circ,y_\circ)$. This is
obtained by imposing $a_{1,k}=0$ and $a_{2,k}=0$.  Since the detector
axes do not necessarily have the same scales nor are perpendicular to
each other, $b_{1,k}$ and $b_{2,k}$ must be free to assume whatever
values fit best.  Therefore, we have to compute in fact only 16
coefficients (for each chip) to derive our GD solution.

Each $i$-star in the master frame is conformally transformed into each
$k$-chip/$j$-image, and cross-identified with the closest source. We
indicate such transformed positions with $(X_i^{\rm
master})^{T_{j,k}}$ and $(Y_i^{\rm master})^{T_{j,k}}$.  Each of such
cross-identifications, when available (of course not all the red
sources in the master list were available in the WFC3 UV-filters),
generates a pair of positional residuals:
$$
\left\{
\begin{array}{rcl}
\displaystyle \Delta x_{i,j,k}&\!\!\!=\!\!\!&x_{i,j,k}^{\rm corr}-(X_i^{\rm master})^{T_{j,k}}\\
\displaystyle \Delta y_{i,j,k}&\!\!\!=\!\!\!&y_{i,j,k}^{\rm corr}-(Y_i^{\rm master})^{T_{j,k}},\\
\end{array}
\right.
$$ which reflect the residuals in the GD (with the opposite sign), and
depend on where the $i$-star fell on the $k/j$-chip/image (plus random
deviations due to non-perfect PSF-fitting and photon noise).  Note
that our calibration process is an iterative procedure, and that
necessarily, at the first iteration, we have to impose $(x^{\rm
corr},y^{\rm corr})_{i,j,k}$ $=$ $(x,y)_{i,j,k}$.  In each chip/image
we have typically $\sim$5500 high-signal unsaturated stars in common
with the master frame, leading to a total of $\sim$50$\,$000 residual
pairs per chip.  [A color-magnitude diagram of the stars actually used
to compute the GD solution is shown, for each filter, in
Fig.\ref{fig:0B}.]

For each chip, these residuals were then collected into a look-up
table made up of $37\times19$ elements, each related to a region of
$110\times110$ pixels.  We chose this particular grid setup because it
offers the best compromise between the need for an adequate number of
grid points to model the GD, and an adequate sampling of each grid
element, containing at least 60 pairs of residuals.  For each grid
element, we computed a set of five 3$\sigma$-clipped quantities:
$\overline{x}_{m,n,k}$, $\overline{y}_{m,n,k}$, $\overline{\Delta
x}_{m,n,k}$, $\overline{\Delta y}_{m,n,k}$, and $P_{m,n,k}$; where
$\overline{x}_{m,n,k}$ and $\overline{y}_{m,n,k}$ are the averaged
positions of all the stars within the grid element $(m,n)$ of the
$k$-chip, $\overline{\Delta x}_{m,n,k}$ and $\overline{\Delta
y}_{m,n,k}$ are the average residuals, and $P_{m,n,k}$ is the number
of stars that were used to calculate the previous quantities. These
$P_{m,n,k}$ will also serve in associating a weight to the grid cells
when we fit the polynomial coefficients.

To obtain the 16 coefficients describing the two polynomials
($a_{q,k}$ with $q=3,\dots,9$, and $b_{q,k}$ with $q=1,\dots,9$),
which represent our GD solution in each chip, we perform a linear
least-square fit of the $N=m\times n= 37 \times 19 = 703$ data points.
Thus, for each chip, we can compute the average distortion correction
in each cell ( $\overline{\delta x}_{p,k}$, $\overline{\delta
y}_{p,k}$ ) with $N$ relations of the form:
$$ 
k=1, 2; ~ p=1,\dots,N: ~ \left\{
\begin{array}{c}
\overline{\delta x}_{p,k}=\displaystyle \sum_{q=3}^9 a_{q,k} t_{q,p,k}\\
\overline{\delta y}_{p,k}=\displaystyle \sum_{q=1}^9 b_{q,k} t_{q,p,k}\\
\end{array}
\right.
$$ (where $t_{1,p,k}= {\overline{\tilde{x}}}_{p,k}$, $t_{2,k} =
{\overline{\tilde{y}}}_{p,k}$, \dots, $t_{9,k} =
{\overline{\tilde{y}^3}}_{p,k}$), and where the 16 unknown quantities
--- $a_{q,k}$ and $b_{q,k}$ --- are our fitting parameters (16 for
each chip) .

In order to solve for $a_{q,k}$ and $b_{q,k}$, we formed, for each
chip, one $9\times9$ matrix $\mathcal M_k$ and two $9\times1$ column
vectors ${\mathcal V}_{a,k}$ and ${\mathcal V}_{b,k}$:
$$
\mathcal{M}_k=\displaystyle  \left(
\begin{array}{cccc}
\sum_p P_{p,k}  t_{1,p,k}^2& \sum_p P_{p,k} t_{1,p,k} t_{2,p,k}  &\cdots& \sum_p P_{p,k}
t_{1,p,k}  t_{9,p,k}\\ \sum_p  P_{p,k} t_{2,p,k}  t_{1,p,k}& \sum_p  P_{p,k} t_{2,p,k}^2
&\cdots&  \sum_p P_{p,k}  t_{2,p,k}  t_{9,p,k}\\ \vdots&\vdots&\ddots&\vdots\\
\sum_p P_{p,k} t_{9,p,k} t_{1,p,k} & \sum_p P_{p,k} t_{9,p,k} t_{2,p,k}&\cdots& \sum_p
P_p t_{9,p,k}^2\\
\end{array}
\right);
$$
$$ 
\mathcal{V}_{a,k}=\left(
\begin{array}{c}
\sum_p  P_{p,k}  t_{1,p,k}   \overline{\Delta  x}_{p,k}\\  \sum_p  P_{p,k}  t_{2,,p,k}
\overline{\Delta x}_{p,k}\\ \vdots  \\ \sum_p P_{p,k} t_{9,p,k} \overline{\Delta
x}_{p,k}\\
\end{array}
\right);
\,\,\,\,\,\,\,\,\,\,\,\,\,\,\,
\mathcal{V}_{b,k}=\left(
\begin{array}{c}
\sum_p P_{p,k} t_{1,p,k} \overline{\Delta y}_{p,k}\\
\sum_p P_{p,k} t_{2,p,k} \overline{\Delta y}_{p,k}\\
\vdots \\
\sum_p P_{p,k} t_{9,p,k} \overline{\Delta y}_{p,k}\\
\end{array}
\right).
$$

The solution is given by two $9\times1$ column vectors $\mathcal A_k$
and $\mathcal B_k$, containing the best fitting values for $a_{q,k}$
and $b_{q,k}$, obtained as:
$$
{\mathcal A_k}=\left(
\begin{array}{c}
a_{1,k}\\
a_{2,k}\\
\vdots\\
a_{9,k}\\
\end{array}
\right)
={\mathcal M_k}^{-1}{\mathcal V}_{a,k};
\,\,\,\,\,\,\,\,\,\,\,\,\,\,\,
{\mathcal B_k}=\left(
\begin{array}{c}
b_{1,k}\\
b_{2,k}\\
\vdots\\
b_{9,k}\\
\end{array}
\right)
={\mathcal M_k}^{-1}{\mathcal V}_{b,k}.
$$

With the first set of calculated coefficients $a_{q,k}$ and $b_{q,k}$
we computed the corrections $\delta x_{i,j,k}$ and $\delta y_{i,j,k}$
to be applied to each $i$-star of the $k$-chip in each $j$-MEF file,
but actually we corrected the positions only by half of the
recommended values, to guarantee convergence.  With the new improved
star positions, we start-over and re-calculated new residuals.  The
procedure is iterated until the difference in the average correction
from one iteration to the following one ---for each grid point---
became smaller than 0.001 pixels.  Convergence was reached after
$\sim100$ iterations.  The coefficients of the final GD solutions for
the two chips, and for the three different filters, are given in
Table~\ref{tab:1}.

\begin{figure*}[t!]
\centering
\includegraphics[width=14cm]{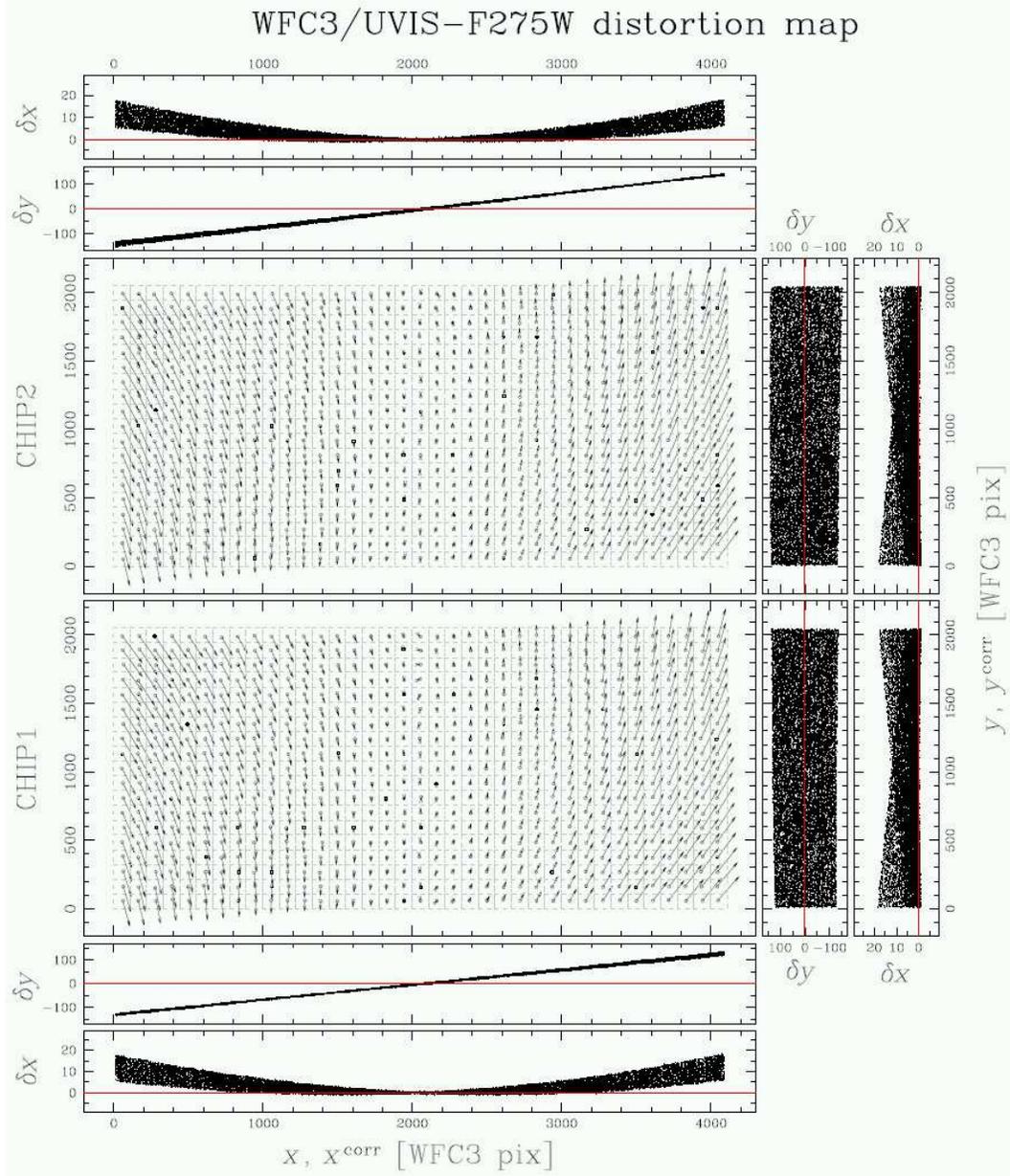}
\caption{Predicted vs.\ uncorrected positions.  The size of the
residual vectors is magnified by a factor of $\times$8 in $x$ and
$\times$1.5 in $y$.  For each chip we plot also individual residuals
as function of $x$ and $y$ axes. Units are expressed WFC3/UVIS pixels
in the reference positions ($x_\circ,y_\circ$). For clarity only a
random 40\% of the residuals is plotted.}
\label{fig1}
\end{figure*}

In Figure~\ref{fig1} we show for the intermediate filter F275W the
total residuals of uncorrected star positions vs.\ the predicted
positions of the master frame, which is representative of our GD
solutions.  For each chip, we plot the $37\times19$ cells used to
model the GD, each with its distortion vector magnified (by a factor
of $\times$8 in $x$, and by a factor of $\times$1.5 in $y$).  Residual
vectors go from the average position of the stars belonging to each
grid cell $(\overline{x},\overline{y})$ to the corrected one
$(\overline{x^{\rm corr}},\overline{y^{\rm corr}})$.  Side panels show
the overall trends of the individual residuals $\delta x$, $\delta y$
along $x$ and $y$ directions (where for clarity we plot only a 40\%
sub-sample, randomly selected).  It immediately strikes the large
linear terms in $y$, reaching up to $\sim$140 pixels.

\begin{figure*}[t!]
\centering
\includegraphics[width=14cm]{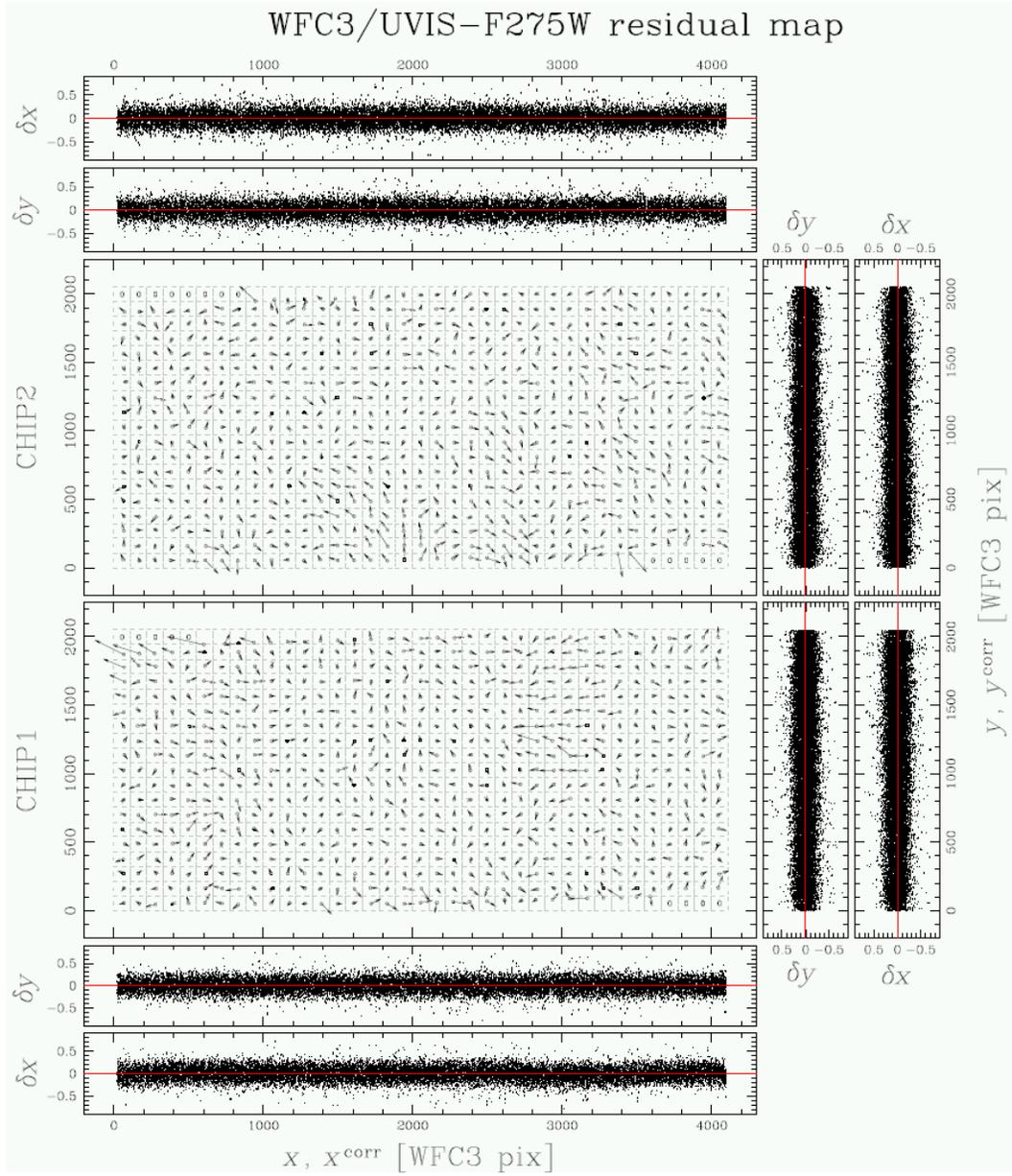}
\caption{Same as Fig.~\ref{fig1} after the correction was applied. The
size of the residuals is now magnified by a factor of 1500.}
\label{fig2}
\end{figure*}

In Figure~\ref{fig2} we show, in the same way, the remaining residuals
after our GD solution is applied.  This time we magnified the
distortion vectors by a factor $\times$1500 in both axes.

At this point it is very interesting to examine the r.m.s.\ of these
remaining residuals, that show a rather large $\sim$0.15 pixels
dispersion.  We will see, in the following, that this dispersion can
be interpreted as the effect of the internal motions of the cluster
stars on the time baseline of $\sim$7 years between the ACS/WFC
observations of the reference frame, and the new WFC3/UVIS data set.
Indeed, assuming 1) a distance of 4.7 kpc for $\omega$~Cen (van der
Marel \& Anderson 2009), 2) an internal velocity dispersion of
$\sim$18 km s$^{-1}$ in our fields (van de Ven et al.\ 2006), and 3)
an isotropic velocity distribution for stars, we would expect to
observe in $\sim$7 years a dispersion of the displacements of
$\sim$5.5 mas.  This dispersion, assuming a pixel scale of $\sim$40
mas for WFC3/UVIS, corresponds to a displacement of $\sim$0.14 pixels
(also in good agreement with the recent measurements by Anderson \&
van der Marel 2009).

\begin{figure*}[t!]
\centering
\includegraphics[width=14cm]{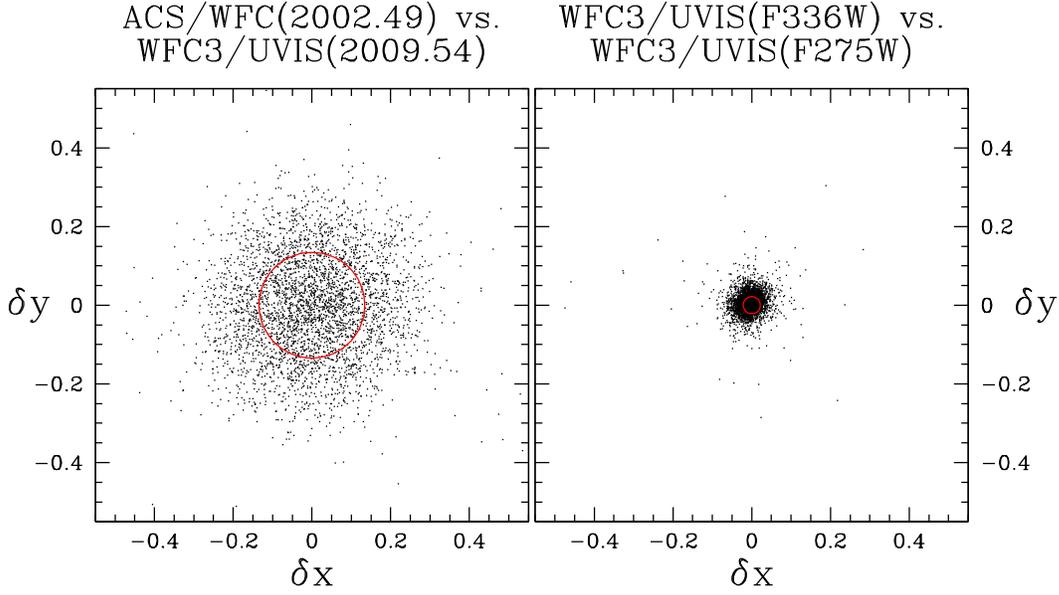}
\caption{ {\it Left:} Vector-point diagram of displacements for the
stars in common between the ACS/WFC epoch of the master catalog, and
the average of the corrected WFC3/UVIS new data in filter F275W.  The
internal motions of $\omega$~Cen dominate the observed dispersions,
but do not prevent a GD solution accurate to $\sim$0.025 WFC3 pixels.
{\it Right:} Vector-point diagram of displacements for the stars in
common between the corrected WFC3/UVIS data in filter F336W, and those
corrected for F275W. The images are collected at the same epoch, and
no sizable internal motions are present.  In this case, the dispersion
reflects our errors.  The red circles, in both panels, indicate the
1-dimensional dispersion of the residuals, and all quantities are
expressed in units of WFC3/UVIS pixels.}
\label{fig:3}
\end{figure*}

To show this more clearly we intercompare the average positions of the
nine WFC3/UVIS corrected catalogs in the filter F275W, with those in
the F336W, for the stars in common between the two filters.  All these
images were taken at the same epoch, and so positions of stars are not
affected by internal motion effects. The 1-dimension dispersion should
reflect our accuracy, and indeed the observed residuals --in this
case-- have a dispersion of $\sim$0.025 pixels (i.e.\ $\sim$1 mas).
Figure~\ref{fig:3} illustrates the two situations.  On the left-panel
we plot the displacements between the ACS/WFC epoch of the reference
frame and the new WFC3/UVIS epoch, while on the right-panel we show
the displacements between our corrected position in filter F275W and
the corrected positions in F336W.  On the left-panel the internal
motions of $\omega$~Cen dominate the dispersion, while on the
right-panel, there are no internal motions at all, and what we are
left with are our errors only.

Unfortunately the WFC3/UVIS images were either not enough, or not well
dithered to perform a pure auto-calibration, and we had to use the
ACS/WFC reference frame.  Nevertheless, even a dispersion of 0.15
pixel within a given cell should be reduced to less than 0.02 pixels
if averaged over more than 60 residuals.  And this should be regarded
as an upper limit, since we are using 703 grid points to constrain 16
parameters.

For this reason, the estimated 0.025 pixel accuracy is larger than we
would have expected.  We can not exclude that these residuals could be
due to a deviation from an isotropic distribution of the internal
motion of $\omega$~Cen (i.e.\ at the level of $\lesssim 3$ km
s$^{-1}$), or simply by unexpectedly large errors in the adopted
ACS/WFC astrometric flat-field (the master frame).  Another
possibility is that there could be some unexpected (and so far
undetected) manufacturing artifact in the WFC3/UVIS detectors which
could affect the positions (such as those identified on WFPC2 CCDs,
and characterized by Anderson \& King 1999, or those of ACS/WFC found
by Anderson 2002).  Finally, it could simply be a higher frequency
spatial variation which can not be properly represented by a
polynomial of a reasonable order, but rather by a residual table as
done in Anderson (2006) for ACS/WFC.  Surely, more data are needed to
further improve the GD solutions presented in this work, as well as a
better time-baseline for the understanding of its variations.  We want
to end this section by pointing out that the detection of the internal
motions among the stars of a Galactic globular cluster is a rather
challenging measurement, and it could well be one of the best
demonstrations of the goodness of our derived geometric distortion
solutions.

\begin{table*}[th!]
\caption{ The coefficients of the third-order polynomial for each chip
and filter.}
\label{tab:1}
\footnotesize{
\centering
\begin{tabular}{ccrrrr}
& & & & & \\ 
\hline
\hline
& & & & & \\
Term $\!(w)\!\!\!\!\!$&Polyn.$\!\!\!\!\!\!\!\!\!$
&$a_{w,[1]}$&$b_{w,[1]}$&$a_{w,[2]}$&$b_{w,[2]}$   \\ 
& & & & & \\ 
\hline
& & & & & \\ 
\multicolumn{6}{c}  {\sc WFC3/UVIS filter F225W} \\
& & & & & \\ 
1&$\tilde{x}$&            $  0.000 \!$&$\! 129.230 \!$&$\!  0.000  \!$&$\! 140.270 \!$\\
2&$\tilde{y}$&            $  0.000 \!$&$\!   1.935 \!$&$\!  0.000  \!$&$\!  -4.215 \!$\\
3&$\tilde{x}^2$&          $ 12.120 \!$&$\!   0.591 \!$&$\! 12.021  \!$&$\!   0.773 \!$\\
4&$\tilde{x}\tilde{y}$&   $ -6.279 \!$&$\!   5.553 \!$&$\! -6.057  \!$&$\!   5.496 \!$\\
5&$\tilde{y}^2$&          $  0.064 \!$&$\!  -3.227 \!$&$\!  0.001  \!$&$\!  -3.058 \!$\\
6&$\tilde{x}^3$&          $  0.176 \!$&$\!   0.029 \!$&$\!  0.149  \!$&$\!   0.156 \!$\\
7&$\tilde{x}^2\tilde{y}$& $ -0.057 \!$&$\!   0.033 \!$&$\!  0.022  \!$&$\!  -0.009 \!$\\
8&$\tilde{x}\tilde{y}^2$& $  0.004 \!$&$\!  -0.041 \!$&$\!  0.061  \!$&$\!  -0.026 \!$\\
9&$\tilde{y}^3$&          $  0.035 \!$&$\!  -0.023 \!$&$\!  0.032  \!$&$\!   0.028 \!$\\
& & & & & \\ 
\multicolumn{6}{c}  {\sc WFC3/UVIS filter F275W} \\
& & & & & \\ 
1&$\tilde{x}$&            $  0.000 \!$&$\! 129.270 \!$&$\!  0.000 \!$&$\! 140.285 \!$\\
2&$\tilde{y}$&            $  0.000 \!$&$\!   1.925 \!$&$\!  0.000 \!$&$\!  -4.221 \!$\\
3&$\tilde{x}^2$&          $ 12.102 \!$&$\!   0.581 \!$&$\! 12.016 \!$&$\!   0.781 \!$\\
4&$\tilde{x}\tilde{y}$&   $ -6.284 \!$&$\!   5.547 \!$&$\! -6.040\!$&$\!   5.493 \!$\\
5&$\tilde{y}^2$&          $  0.061 \!$&$\!  -3.241 \!$&$\!  0.001 \!$&$\!  -3.048 \!$\\
6&$\tilde{x}^3$&          $  0.178 \!$&$\!   0.033 \!$&$\!  0.144 \!$&$\!   0.163 \!$\\
7&$\tilde{x}^2\tilde{y}$& $ -0.056 \!$&$\!   0.054 \!$&$\!  0.026 \!$&$\!   0.007 \!$\\
8&$\tilde{x}\tilde{y}^2$& $  0.005 \!$&$\!  -0.041 \!$&$\!  0.051 \!$&$\!  -0.025 \!$\\
9&$\tilde{y}^3$&          $  0.033 \!$&$\!  -0.012 \!$&$\!  0.032 \!$&$\!   0.020 \!$\\
& & & & & \\ 
\multicolumn{6}{c}  {\sc WFC3/UVIS filter F336W} \\
& & & & & \\ 
1&$\tilde{x}$&            $  0.000 \!$&$\! 129.438 \!$&$\!  0.000  \!$&$\!  140.315 \!$\\
2&$\tilde{y}$&            $  0.000 \!$&$\!   1.786 \!$&$\!  0.000  \!$&$\!   -4.322 \!$\\
3&$\tilde{x}^2$&          $ 12.091 \!$&$\!   0.676 \!$&$\! 11.994  \!$&$\!    0.672 \!$\\
4&$\tilde{x}\tilde{y}$&   $ -6.188 \!$&$\!   5.565 \!$&$\! -6.135  \!$&$\!    5.476 \!$\\
5&$\tilde{y}^2$&          $  0.065 \!$&$\!  -3.155 \!$&$\!  0.004  \!$&$\!   -3.152 \!$\\
6&$\tilde{x}^3$&          $ -0.062 \!$&$\!   0.004 \!$&$\! -0.151  \!$&$\!    0.189 \!$\\
7&$\tilde{x}^2\tilde{y}$& $ -0.097 \!$&$\!   0.034 \!$&$\!  0.074  \!$&$\!   -0.027 \!$\\
8&$\tilde{x}\tilde{y}^2$& $  0.016 \!$&$\!  -0.061 \!$&$\!  0.040  \!$&$\!    0.005 \!$\\
9&$\tilde{y}^3$&          $  0.033 \!$&$\!   0.016 \!$&$\!  0.033  \!$&$\!    0.014 \!$\\
& & & & & \\ 
\hline
\hline
\end{tabular}}
\end{table*}

~\\
\newpage
~\\
\newpage
~\\
\newpage
~\\
\newpage

%
% alpha = 1.005949277 +/- 2.016203168e-05
% theta = 0.06539555556 +/- 0.0009324753382
% xo = 2046.004267 +/- 0.03014009364
% yo = 3098.336204 +/- 0.03053958183
%
\begin{table*}[th!]
\caption{Interchip transformation parameters. Chip [1] parameters are
indicated only for clarity. For chip [2] formal errors are given.}
\label{tab:2}
\centering
\begin{tabular}{ccccc}
 & & & & \\
\hline 
\hline 
 & & & & \\
$k$-chip  & $\alpha_{[k]}/\alpha_{[1]}$ & $\theta_{[k]}$$-$$\theta_{[1]}$ & 
  $(x_{\circ}^{[k]})^{\rm corr}_{[1]}$ & $(y_{\circ}^{[k]})^{\rm corr}_{[1]}$ \\
 & & & & \\ 
       & $[$number$]$    & $[^\circ]$ & $[$pixel$]$ & $[$pixel$]$ \\
 & & & & \\ 
\hline  
 & & & & \\ 
$[1]$ & 1.00000        &   0.0000     &   2048.00 &   1025.00 \\ 
 & & & & \\ 
$[2]$ & 1.00595        &   0.0654     &   2046.00 &   3098.34 \\ 
      & $\pm$2/$100\,000$ & $\pm$0.001 & $\pm$0.03 & $\pm$0.03 \\ 
 & & & & \\ 
\hline
\hline
\end{tabular}
\end{table*}

\section{Interchip transformations}

\begin{figure*}[ht!]
\centering
\includegraphics[width=8cm]{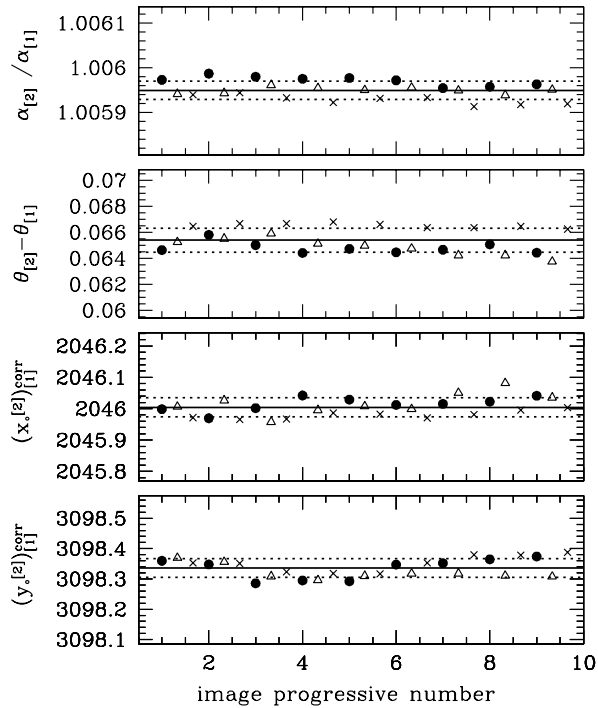}
\caption{Interchip transformation parameters as obtained from
individual images.  Data points from F225W are indicated with
filled-circles, F275W with triangles, and F336W with crosses. The
averages are indicated with solid lines, while the dashed-lines give
the formal uncertainties.}
\label{fig:4}
\vskip 2mm
\end{figure*}

For many applications it would be useful to transform the GD corrected
positions of each chip into a common distortion-free reference frame.
We could then, simply conformally transform the corrected positions of
chip [k] into the distortion corrected positions of chip~[1], using
the following relations:\
$$
\left(
\begin{array}{c}
x^{\rm corr}_{[1]}\\
y^{\rm corr}_{[1]}\\
\end{array}
\right)
= ~ \frac{\alpha_{[1]}}{\alpha_{[k]}} ~  
\left[
\begin{array}{cc}
 \cos(\theta_{[1]}-\theta_{[k]}) & \sin(\theta_{[1]}-\theta_{[k]}) \\
-\sin(\theta_{[1]}-\theta_{[k]}) & \cos(\theta_{[1]}-\theta_{[k]}) \\
\end{array}
\right]
\left(
\begin{array}{c}
x_{[k]}^{\rm corr} - 2048 \\
y_{[k]}^{\rm corr} - 1025 \\
\end{array}
\right)
+
\left(
\begin{array}{c}
(x_{\circ}^{[k]})^{\rm corr}_{[1]} \\
(y_{\circ}^{[k]})^{\rm corr}_{[1]} \\
\end{array}
\right);  
$$ where ---following the formalism in AK03--- we indicate the scale
factor as as $\alpha_{[k]}$, the orientation angle with
$\theta_{[k]}$, and the positions of the center of the chip
$(x_\circ,y_\circ)$ in the corrected reference system of chip [1] as
$(x_{\circ}^{[k]})^{\rm corr}_{[1]}$ and $(y_{\circ}^{[k]})^{\rm
corr}_{[1]}$.  Of course, for $k=1$, we end up with the identity.  The
values of the interchip transformation parameters are given in
Table~\ref{tab:2}, and shown for individual images in
Figure~\ref{fig:4}.

~\\
~\\

~\\
\newpage 

\section{Average Absolute Scale relative to ACS/WFC}

\begin{figure*}[ht!]
\centering
\includegraphics[width=8cm]{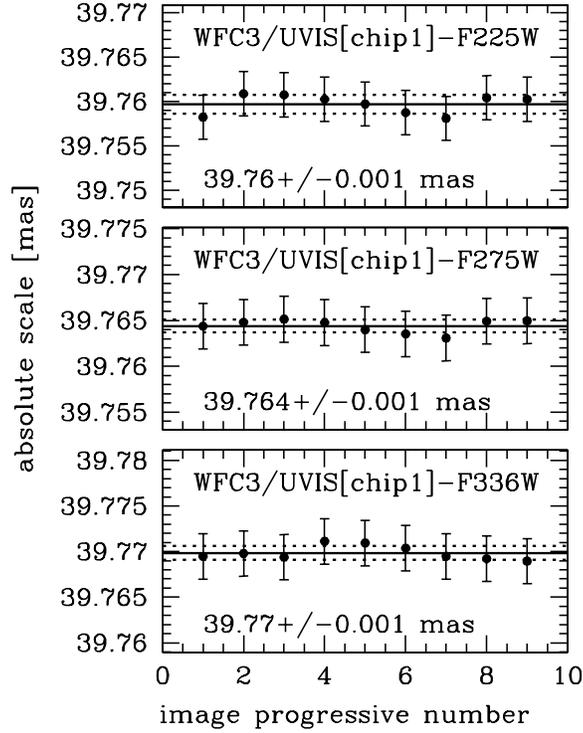}
\caption{Absolute scales relative to the one adopted for our ACS/WFC
master frame expressed in mas.  Each panel shows individual images,
for different filters.  As in the previous figure, the average values
are indicated with solid lines, while the dashed-lines give the formal
uncertainties.  For reference, we show with error-bars the on-orbit
maximum systematic errors introduced by the velocity aberration on the
plate scale, i.e.\ for a velocity of $\pm$7 km s$^{-1}$.  However,
since our master frame has not been corrected for scale variation
induced by velocity aberration, these values can not be considered
more accurate --in absolute sense-- than 12 parts in $100\,000$.}
\label{fig:5}
\vskip 2mm
\end{figure*}

The final step is to link, for each filter, the WFC3/UVIS chip [1] to
an absolute plate scale in mas.  To this purpose we adopt an average
plate scale for our ACS/WFC master frame of 49.7248 mas
ACS/WFC-pixel$^{-1}$ (from van der Marel et al.\ 2007), and multiplied
it by the --measured-- scale factor between the WFC3/UVIS chip [1] and
the master frame (which is expressed in ACS/WFC pixels).  The results
for the individual images and the averages for each filter, are shown
in Figure~\ref{fig:5}, while Table~\ref{tab:3} gives the average
values in mas pixel$^{-1}$. We believe that the differences in the
relative values for the three filters are significant. The fact that
the plate scales correlate with the wavelength suggests that
refraction introduced by either the filters, or by the two
fused-silica windows of the dewar, could have some role.

Concerning their absolute values, instead, we have to consider
that the velocity of {\it HST} around the Earth ($\pm$7 km s$^{-1}$)
causes light aberration which induces plate-scale variations up to 5
parts in $100\,000$ (Cox \& Gilliland 2002)\footnote{If we sum to
this the Earth velocity around the Sun the plate-scale variations can
reach up to 12 parts in $100\,000$ (Cox \& Gilliland 2002).}, and
that our master frame (from Villanova et al.\ 2007) was not corrected
for it.

The ACS/WFC plate scale for the Anderson's (2006, 2007) GD solution
--once corrected for the temporal variations of the linear terms-- has
proved to be stable at a level of accuracy better than these velocity
aberration variations (van der Marel et al.\ 2007).  However, since we
are not attempting to correct for this effect on our adopted ACS/WFC
master frame, we simply limit the accuracy of the here derived
WFC3/UVIS plate-scale absolute values to these accuracies, i.e.\ 12
parts in $100\,000$.

\begin{table*}[th!]
\caption{Average absolute plate scale of WFC3/UVIS in mas
pixel$^{-1}$.  Accuracy is 12 parts in $100\,000$.}
\label{tab:3}
\centering
\begin{tabular}{cccc}
  & & & \\
\hline 
\hline 
  & & & \\
 $[$mas pixel$^{-1}$$]$ & F225W  & F275W  & F336W \\ 
  & & & \\ 
\hline  
  & & & \\ 
$\alpha_{[1]}$ & 39.760 & 39.764 & 39.770 \\  
  & & & \\ 
\hline
\hline
\end{tabular}
\end{table*}

~\\
\newpage
~\\
\newpage

\section{Conclusions}
\label{sec:3}

By using a limited (but best available) number of exposures with large
dithers, and an existing ACS/WFC astrometric flat field, we have found
a set of third-order correction coefficients to represent the
geometric distortion of WFC3/UVIS in three broad-band ultraviolet
filters. The solution was derived independently for each of its two
CCDs.

The use of these corrections removes the distortion over the entire
area of each chip to an average accuracy of $\sim$0.025 pixel (i.e.\
$\sim$1 mas), the largest systematics being located in the $\sim$200
pixels closest to the boundaries of the detectors (and never exceeding
0.06 pixels).  We advise the use of the inner parts of the detectors
for high-precision astrometry.  The limitation that has prevented us
from removing the distortion at an even higher level of accuracy is
the lack of enough observations collected at different roll-angles and
dithers which could enable us to perform an auto-calibration.

Nevertheless, the comparison of the mid-2002 ACS/WFC positions with
the new WFC3 observations corrected with our astrometric solutions are
good enough to clearly show the internal motions of $\omega$~Centauri.
These proved to be in perfect agreement with the most recent
determinations.

We also derived the average absolute scale of the detector with an
accuracy limited by the uncertainties in the plate-scale variations
induced by the velocity aberration of the telescope motion in the
Earth-Sun system.

For the future, more data with a longer time-baseline are needed to
better characterize the GD stability of {\it HST} WFC3/UVIS detectors
in the medium and long term.

\acknowledgements We thank an anonymous (and very competent) referee
for a careful reading, and for the useful corrections/suggestions.  We
thank Giampaolo Piotto for reading the manuscript, ``Alcicci'' Z.\
Bonanos for polishing it, and George Hartig for a useful discussion.
We thank Jay Anderson for having trained us (for almost ten years) in
his astrometric {\it arts}, without which this work would have not
been possible.  A.B.\ acknowledges the support by the CA.RI.PA.RO.\
foundation, and by the STScI under the \textit{2008 graduate research
assistantship} program.  Finally, we thank {\sf FORTRAN-77}, {\sf
LINUX}, and {\sf SuperMongo} to make this work easy.

\end{document}